\documentclass[prd,tightenlines,twoside,secnumarabic,
               onecolumn,floatfix,nofootinbib,showpacs,11pt]{revtex4}

\usepackage[english]{babel}
\usepackage{amsmath,amssymb,bm,slashed}
\usepackage{graphicx}
\usepackage[sort&compress]{natbib}
\usepackage{xcolor}
\usepackage[normalem]{ulem}
\usepackage{hyperref}
\usepackage{cleveref}
\usepackage{subfigure} 
\definecolor{red}{rgb}{1.0, 0, 0}

\allowdisplaybreaks

\bibpunct{[}{]}{,}{n}{}{,}

\newcommand{\ket}[1]{\ensuremath{| #1 \rangle}}   
\newcommand{\ev}[1]{\ensuremath{\left\langle #1 %
                     \right\rangle}} 

\newcommand{\diag}{\text{diag}}

\renewcommand{\vec}[1]{{\mathbf{#1}}}

\providecommand{\abs}[1]{\lvert#1\rvert}

\DeclareMathOperator{\Lagr}{\mathcal{L}}

\begin{document}

\title{The Not-So-Sterile 4th Neutrino: \\
Constraints on New Gauge Interactions from Neutrino Oscillation Experiments}
\author{Joachim Kopp$^{1,2}$}    \email[Email: ]{jkopp@uni-mainz.de}
\author{Johannes Welter$^{1}$}   \email[Email: ]{welter@mpi-hd.mpg.de}
\affiliation{
  $^1$ Max Planck Institut f\"ur Kernphysik, Saupfercheckweg 1, 69117 Heidelberg, Germany \\
  $^2$ PRISMA Cluster of Excellence and Mainz Institute for Theoretical Physics,
    Johannes Gutenberg University, 55099 Mainz, Germany}
\date{\today} 

\begin{abstract}
  Sterile neutrino models with new gauge interactions in the sterile sector are
  phenomenologically interesting since they can lead to novel effects in neutrino oscillation
  experiments, in cosmology and in dark matter detectors, possibly
  even explaining some of the observed anomalies in these experiments.  Here,
  we use data from neutrino oscillation experiments, in particular from
  MiniBooNE, MINOS and solar neutrino experiments, to constrain such models.
  We focus in particular on the case where the sterile sector gauge boson $A'$
  couples also to Standard Model particles (for instance to the baryon number
  current) and thus induces a large Mikheyev-Smirnov-Wolfenstein potential.
  For eV-scale sterile neutrinos, we obtain strong constraints especially from
  MINOS, which restricts the strength of the new interaction to be less than
  $\sim 10$ times that of the Standard Model weak interaction unless
  active--sterile neutrino mixing is very small ($\sin^2 \theta_{24} \lesssim
  10^{-3}$).  This rules out gauge forces large enough to affect short-baseline
  experiments like MiniBooNE and it imposes nontrivial constraints on signals from
  sterile neutrino scattering in dark matter experiments.
\end{abstract}

\begin{flushright}
  MITP/14-056
\end{flushright}

\maketitle


\section{Introduction and motivation}
\label{sec:intro}

The possible existence of sterile neutrinos (Standard Model singlet fermions)
with masses of order eV has been a widely discussed topic in astroparticle
physics over the past few years. It is motivated by several anomalous results
from short-baseline neutrino oscillation experiments, in particular the
excesses of $\nu_e$ and $\bar\nu_e$ events in a $\nu_\mu$ and $\bar\nu_\mu$ beam respectively observed by
LSND~\cite{Aguilar:2001ty} and MiniBooNE~\cite{AguilarArevalo:2012va}, the
apparently lower than expected $\bar\nu_e$ flux from nuclear
reactors~\cite{Mueller:2011nm, Mention:2011rk, Huber:2011wv} (see however
\cite{Hayes:2013wra}) and the deficit of $\nu_e$ in radioactive source
experiments~\cite{Acero:2007su,Giunti:2010zu}.  Global fits~\cite{Kopp:2011qd,
Giunti:2011cp, Karagiorgi:2011ut, Giunti:2011hn, Giunti:2011gz,
Abazajian:2012ys, Karagiorgi:2011ut, Kopp:2013vaa} show that these anomalies
could be explained if sterile neutrinos with $\mathcal{O}(\text{eV})$ mass and
$\mathcal{O}(10\%)$ mixing with $\nu_e$ and $\nu_\mu$ exist.  However, global
fits also reveal that it is difficult to reconcile such a scenario with
existing null results from other short-baseline oscillation experiments.

Constraints come also from cosmological observations, which slightly disfavor
scenarios with extra relativistic degrees of freedom in the early
Universe~\cite{Ade:2013zuv}. Cosmology also imposes a tight constraint on the
sum of neutrino masses $\sum_j m_{\nu j} < 0.23$, where the sum runs over all
neutrino mass eigenstates that are in thermal equilibrium in the early
Universe.  Note that these constraints would be relaxed if the recent BICEP-2
data on B-modes in the cosmic microwave background~\cite{Ade:2014xna} is
confirmed~\cite{Giusarma:2014zza, Dvorkin:2014lea, Li:2014cka, Zhang:2014dxk,
Archidiacono:2014apa}.

An interesting scenario that is unconstrained by cosmology is
\emph{self-interacting sterile neutrinos}~\cite{Hannestad:2013ana,
Dasgupta:2013zpn}.  If interactions among sterile neutrinos are mediated by a
scalar or gauge boson with a mass of order MeV or lighter, sterile neutrinos
will feel a strong thermal potential in the early Universe which suppresses
their mixing with active neutrinos and thus prohibits their production through
oscillations. Moreover, if the new interaction couples not only to sterile
neutrinos, but also to dark matter, it has the potential to explain several
problems with cosmic structure formation at small
scales~\cite{Dasgupta:2013zpn, Bringmann:2013vra}.

If a new interaction is shared between sterile neutrinos and ordinary matter
(for instance in models with gauged baryon number coupled to sterile neutrinos
and in scenarios in which a sterile sector gauge boson mixes kinetically with
the photon), interesting signals in direct dark matter searches are
possible~\cite{Pospelov:2011ha, Harnik:2012ni, Pospelov:2012gm,
Pospelov:2013rha}. The increased neutrino--nucleus scattering cross section
might even explain some of the excess events observed by several experiments.
On the other hand, such scenarios are more challenging for cosmology because of
an additional sterile neutrino production mechanism through the gauge
interaction.  (Note that these constraints are still avoided for instance in
scenarios with extra entropy production in the visible sector after sterile
neutrino decoupling~\cite{Ho:2012br}.)

In this paper, we investigate how novel interactions between sterile neutrinos
and ordinary matter are constrained by neutrino oscillation experiments at
short and long baseline. This topic has been discussed in a previous
paper~\cite{Karagiorgi:2012kw}, the conclusions of which we will update below.
For definiteness, we will focus on scenarios similar to the
``baryonic sterile neutrino'' scenario first introduced in
\cite{Pospelov:2011ha}, where the sterile neutrino couples to gauged baryon
number. We emphasize, however, that our results are directly applicable to any
theory in which sterile neutrinos interact with Standard Model (SM) fermions through a new gauge
force under which ordinary matter carries a net charge. (The last condition
excludes models in which the coupling is only through kinetic mixing between
the new gauge boson and the photon.)  The new gauge current creates a
Mikheyev-Smirnov-Wolfenstein (MSW) potential for sterile neutrinos propagating
through ordinary matter and has thus a potentially large impact on neutrino
oscillations.  Since the mass of the new gauge boson in this model can be as
low as 10~MeV (see \cite{Harnik:2012ni} for detailed constraints) and since
constraints on its coupling are weak~\cite{Pospelov:2011ha}, the strength of
the effective interaction can be more than two orders of magnitude larger than
the SM weak interactions responsible for the ordinary MSW effect.
This implies that resonant enhancement of the oscillation amplitude could be
relevant at $\mathcal{O}(\text{GeV})$ energies even for relatively large mass
squared difference $\Delta m_{41}^2 \sim \text{eV}$ between the mostly sterile and mostly
active mass eigenstates.  The model could thus potentially allow an explanation
of some of the short-baseline oscillation anomalies with significantly smaller
vacuum mixing angles than in sterile neutrino scenarios without new
interactions.

The structure of the paper is as follows.  In section~\ref{sec:models}, we
briefly review models with new interactions in the sterile sector in general,
and the ``baryonic neutrino'' model from~\cite{Pospelov:2011ha} in particular.
We map these models onto an effective field theory and discuss their implications
for neutrino oscillations. In particular, we derive approximate analytical
formulas for the oscillation probabilities. In
section~\ref{sec:osc-constraints}, we then present our main numerical results,
which will set strong constraints on new forces coupling sterile neutrinos to
SM particles.  We will summarize and conclude in section~\ref{sec:conclusions}.

\section{Models and formalism}
\label{sec:models}

\subsection{New gauge bosons in the sterile neutrino sector}

In the following we shortly describe the model proposed
in~\cite{Pospelov:2011ha,Harnik:2012ni}, originally introduced to
study the impact of a new gauge force in the sterile neutrino sector on dark
matter searches. The basic idea is to introduce a fourth left-handed neutrino
flavour $\nu_b$, sterile under SM interactions, which can have a relatively
large coupling to baryons ($10^2$--$10^3$ times larger than the Fermi constant $G_F$)
without being in conflict with current experimental bounds, like for examples
constraints coming from meson decays such as $K \rightarrow \pi
\bar{\nu}_b \nu_b$~\cite{Pospelov:2011ha}.  It can be implemented by introducing a
new $U(1)_B$ gauge symmetry under which quarks have charge $g_b/3$ and the
baryonic neutrino $\nu_b$ has charge $g_b'$. We will assume $g_b$ and $g_b'$ to
be of order 0.1--1. To cancel anomalies, the introduction of additional fermions
charged under $U(1)_B$ will be necessary, but we assume that these do not mix significantly
with SM neutrinos and can be neglected. The baryonic gauge boson $X$ acquires a mass
when $U(1)_B$ is broken by a new sterile sector Higgs field $h_b$.  The
relevant part of the Lagrangian after symmetry breaking can be written as~\cite{Pospelov:2011ha} 
\begin{eqnarray}
 \label{eqn:Lagr}
 \Lagr &\supset&
           - \frac{1}{4} F_{X,\mu\nu} F_X^{\mu\nu}
           + \frac{1}{2} m_X^2 X_{\mu} X^{\mu} \nonumber \\
       & & + \bar{\nu}_b \gamma_{\mu} \big( i \partial^{\mu} + g_b' X^{\mu} \big) \nu_b
           + \sum_q \bar{q} \Big( i \slashed{D}_{\text{SM}} 
                                   + \frac{1}{3} g_b \gamma_{\mu} X^{\mu} \Big) q
           + \Lagr_m \text{,}
\end{eqnarray}
where $q$ are the SM quark fields, $F_{X,\mu\nu} \equiv \partial_\mu X_\nu - \partial_\nu X_\mu$
is the field strength tensor of the baryonic vector boson $X_\mu$ and $m_X \sim 
1 \, \text{GeV}$ is its mass. In a seesaw framework, the baryonic neutrino mixes with the SM
through the terms
\begin{eqnarray}
 \Lagr_{m} = - \sum_{\alpha,j} m_D^{\alpha j} \bar{\nu}_L^\alpha N_R^j
             - \sum_j m_b^j \bar{\nu}_{bL} N_R^j
             - \frac{1}{2} \sum_{i\text{,}j} m_R^{ij} \overline{\left( N_R^i \right)^C} N_R^j 
             + h.c. \,,
  \label{eqn:Lagr_m}
\end{eqnarray}
with the Dirac mass matrix $m_D$ of the active neutrinos $\nu_L^\alpha$, the
Dirac mass vector of the baryonic neutrino $m_b^j$ and the the Majorana mass
matrix $m_R^{ij}$ of the heavy right-handed neutrino fields $N_R^j$.  The
flavour index $\alpha$ runs over $e$, $\mu$ and $\tau$, while the indices $i$
and $j$ run over all heavy right-handed neutrino states.

The Lagrangian of equation~\eqref{eqn:Lagr} implies the existence of a new MSW potential that sterile 
neutrinos experience while propagating in matter. This effect is caused by 
coherent elastic forward scattering on neutrons and protons and can lead to resonant 
enhancement of flavour oscillations.  Since coherent forward scattering does not
involve any momentum transfer, its amplitude can be most easily obtained from the
low energy effective Lagrangian of baryonic neutral current interactions
\begin{eqnarray}
  \Lagr_{b,\text{eff}} &=&
  \frac{G_B}{2} \big[ \bar{\nu}_b \gamma_{\mu} \left( 1- \gamma_5 \right) \nu_b \big]
                \big[ \bar{p} \gamma^{\mu} p + \bar{n} \gamma^{\mu} n \big] \text{.}
\end{eqnarray}
Here, the effective coupling constant is $G_B \equiv g_b g_b' / m_X^2$.
By treating neutrons and protons as a static background field~\cite{Akhmedov:1999uz},
we obtain the matter potential for sterile neutrinos
\begin{eqnarray}
  V_b = G_B N_{\text{nucl}} \text{.}
\end{eqnarray}
The potential for sterile anti-neutrinos has opposite sign.  Here,
$N_{\text{nucl}}$ is the number density of nucleons in the background matter.
Note that $G_B$ can be either positive or negative, depending on the relative
sign of $g_b$ and $g_b'$.  In the following analysis we will use the ratio of
the coupling constants 
\begin{eqnarray}
  \epsilon \equiv \frac{G_B}{\sqrt{2}G_F}
\end{eqnarray}
as a measure for the relative strength of
$V_b$ compared to the potential $V_{\text{CC}}$ that charged current (CC)
interactions with electrons induce for electron neutrinos in the SM.  The
baryonic potential can be written as
\begin{align}
  V_b &= V_{\text{CC}} \cdot G_B/(\sqrt{2} G_F Y_e) = \epsilon \, V_\text{CC} / Y_e  \\
      &= \epsilon \cdot 7.56 \cdot 10^{-14}\ \text{eV} \cdot 
         \bigg( \frac{\rho}{\text{g}/\text{cm}^3} \bigg),
\end{align}  
where $Y_e$ is the number of electrons per nucleon.

As mentioned in the introduction, baryonic sterile neutrinos could lead to
novel signals in direct dark matter searches thanks to an enhanced sterile
neutrino--nucleus scattering rate.  Typically, observable effects in current
experiments are expected if $\epsilon \gtrsim 100$~\cite{Pospelov:2011ha,
Harnik:2012ni, Pospelov:2012gm, Pospelov:2013rha}. We will see in
section~\ref{sec:results} that such large values of $\epsilon$ are largely
excluded for eV scale sterile neutrinos with substantial mixing into the active
sector.

We wish to stress here that, while we use baryonic sterile neutrinos as a benchmark
scenario, our results will apply to any scenario in which sterile neutrinos
have new gauge interactions with SM fermions.  It is important to keep in mind,
though, that models with new forces in the lepton sector are much more tightly
constrained than new baryonic interactions (see e.g.\ \cite{Harnik:2012ni} for a
review).

The mass terms in equation \eqref{eqn:Lagr_m} lead to flavour mixing between
$\nu_b$ and the active neutrinos, as can be seen by integrating out the
heavy right-handed neutrinos and diagonalizing the resulting mass matrix. In
this way, we obtain the $4 \times 4$ mixing matrix $U$ connecting mass
eigenstates $\ket{\nu_i}$ and flavour eigenstates $\ket{\nu_{\alpha}}$:
\begin{align}
  \ket{\nu_\alpha} = \sum_i U_{\alpha i}^* \ket{\nu_{i}} \text{.}
\end{align}
Since $U$ is unitary, it can be parametrized by $6$ rotation angles
$\theta_{ij}$ and $3$ complex phases $\delta_{ij}$~\footnote{We omit the 
Majorana phases here since they do not contribute to neutrino flavour 
oscillations.}
\begin{eqnarray}
  \label{eqn:parametrization}
  U &=& R_{34} \cdot R_{24}' \cdot R_{14}' \cdot R_{23} \cdot R_{13}' \cdot R_{12} \text{.}
\end{eqnarray}
Here, $R_{ij}$ describes a rotation matrix in the $ij$ plane, while
$R_{ij}'$ corresponds to a complex rotation by the angle $\theta_{ij}$ and
phase $\delta_{ij}$. Given the mixing matrix $U$ and the mass squared difference
$\Delta m^2_{41}$ between the mostly sterile mass eigenstate $\nu_4$ and the mostly active
mass eigenstate $\nu_1$, one can write down the effective
Hamiltonian\footnote{\textit{Effective} means that terms proportional to the
unit matrix are omitted because they do not contribute to flavour oscillations. Also
note that we assume a definite three-momentum that is the same for all
contributing mass eigenstates so that one can approximate $E_i \approx
\abs{\vec{p}} + m_i^2/(2E)$. It is well-known that this approximation, though technically
unjustified, leads to correct results for neutrino oscillation probabilities~\cite{Giunti}.}
in flavour space: 
\begin{eqnarray}
  H^{\text{flavour}}_{\text{eff}} = \frac{1}{2 E} U 
    \begin{pmatrix} 0 &   &   &    \\ 
                      & \Delta m_{21}^2 &   &    \\
                      &   & \Delta m_{31}^2 &    \\
                      &   &   & \Delta m_{41}^2
    \end{pmatrix} U^{\dag}
  + \begin{pmatrix} V_{\text{CC}} &  &  &  \\ 
                      & 0 &   &   \\
                      &   & 0 &   \\
                      &   &   & V_b - V_{NC}
    \end{pmatrix} \text{.} \label{eqn:effectiveH}
\end{eqnarray}
Here, $V_{NC} \equiv -\sqrt{2} G_F n_n / 2$ is the contribution from SM neutral current
interactions to the MSW potential. It is proportional to the number density $n_n$
of neutrons in the background material.

The oscillation probability $P_{\nu_{\alpha} \rightarrow \nu_{\beta}} (t)$,
i.e.~the probability for a neutrino of initial flavour $\alpha$ to be converted
into flavour $\beta$ after traveling a time $t$, can then be obtained by
diagonalizing the effective Hamiltonian according to
$H^{\text{flavour}}_{\text{eff}}=\tilde{U} \diag(\lambda_1, \lambda_2,
\lambda_3, \lambda_4) \tilde{U}^{\dag}$ and inserting the eigenvalues
$\lambda_i$ and the effective mixing matrix $\tilde{U}$ into the well-known
formula 
\begin{equation}
  P_{\nu_{\alpha} \rightarrow \nu_{\beta}}
    = \big| \ev{\nu_{\beta} | \nu_{\alpha}(t)} \big|^2
    = \Big| \sum_j {\tilde{U}_{\alpha j}}^* \tilde{U}_{\beta j} e^{-i \lambda_j L}\Big|^2 \text{.}
  \label{eqn:OscProb}
\end{equation}

\subsection{Approximate oscillation probabilities}
\label{subsec:oscprob}

As a prelude to the numerical fits we are going to present in
section~\ref{sec:osc-constraints}, we give here approximate analytic
expressions for the oscillation probabilities in the baryonic sterile neutrino
model and in models with new sterile neutrino--SM interactions in general.
Similar calculations have been carried out previously
in~\cite{Karagiorgi:2012kw} and we will compare these results to ours in
section~\ref{subsec:ksc}.

Our starting point is to assume $\abs{\Delta m_{41}^2} \gg \abs{\Delta
m_{31}^2}, \Delta m_{21}^2$, which is a good approximation at sufficiently
short baselines.  Moreover, we neglect the SM MSW potentials $V_{\text{NC}}$
(arising from $Z$ exchange diagrams) and $V_{\text{CC}}$ (arising from $W$
exchange diagrams) against the baryonic potential $V_b$, which we assume to be
much larger.  With
these approximations, mixing among the three active flavour eigenstates becomes
irrelevant. (They can, however, still oscillate into each other through their mixing
with $\nu_b$.)  We also set 
$U_{\tau 4}=0$ for simplicity, following~\cite{Karagiorgi:2012kw}. With these assumptions,
diagonalization of the Hamiltonian $H^{\text{flavour}}_{\text{eff}}$ from
equation~\eqref{eqn:effectiveH} yields for the eigenvalues $\lambda_i$
\begin{align}
  \lambda_1 = \lambda_2 = 0 \text{,} \qquad 
  \lambda_3 = \frac{1}{2} \Big( V_b + \frac{\Delta m_{41}^2}{2 E} - A \Big) \text{,} \qquad
  \lambda_4 = \frac{1}{2} \Big( V_b + \frac{\Delta m_{41}^2}{2 E} + A \Big) \text{.}
                                                                          \label{eqn:eigenvalues}
\end{align}
The elements of the unitary matrix $\tilde{U}$ are
\begin{gather}
  \tilde{U}_{\mu 1} = \tilde{U}_{e 1} = 0 \text{,} \qquad
  \abs{\tilde{U}_{e 2}}^2 = \frac{\abs{U_{\mu 4}}^2}{1-\abs{U_{s4}}^2} \text{,} \qquad
  \abs{\tilde{U}_{\mu 2}}^2 = \frac{\abs{U_{e4}}^2}{1-\abs{U_{s4}}^2} \text{,} \nonumber \\[0.4cm]
  \abs{\tilde{U}_{e 4}}^2 =
    \abs{U_{e 4}}^2 \frac{ \frac{\Delta m_{41}^2}{2 E} 
                           \Big[ A+\frac{\Delta m_{41}^2}{2 E}-V_b \Big]}
                         { A \Big[ A+\frac{\Delta m_{41}^2}{2 E}+V_b \Big]} \text{,} \qquad
  \abs{\tilde{U}_{\mu 4}}^2 =
    \abs{U_{\mu 4}}^2 \frac{ \frac{\Delta m_{41}^2}{2 E} 
                             \Big[ A+\frac{\Delta m_{41}^2}{2 E}-V_b \Big]}
                           { A \Big[ A+\frac{\Delta m_{41}^2}{2 E}+V_b \Big]} \text{.}
\end{gather}
Here, we have introduced the abbreviation 
\begin{eqnarray}
  A &=& |V_b| \cdot \sqrt{1+\left( 4 \abs{U_{s4}}^2-2 \right) \frac{\Delta m_{41}^2}{2 E V_b}
                         +\left(\frac{\Delta m_{41}^2}{2 E V_b}\right)^2}                       \text{.}
  \label{eqn:A}
\end{eqnarray}
With these formulas at hand and using the unitarity condition $\sum_i
\tilde{U}_{\alpha i}^* \tilde{U}_{\alpha i} = 1$ as well as the observation that
$\tilde{U}_{\mu 2} \tilde{U}_{e 2}^* \tilde{U}_{\mu 4}^* \tilde{U}_{e 4}$ is real,
it is straightforward to
calculate the oscillation probabilities according to equation~\eqref{eqn:OscProb}. 
For $\alpha = \mu$ and $\beta = \mu$, $e$ we obtain 
\begin{eqnarray}
 P_{\nu_{\mu} \rightarrow \nu_{e}}
    &=&   - 4 \frac{\abs{U_{e4}}^2\abs{U_{\mu 4}}^2\abs{U_{s4}}^2}{1-\abs{U_{s4}}^2} 
              \left(\frac{\Delta m_{41}^2}{2 E A}\right)^2 \sin^2\phi_1
       + 2 \frac{\abs{U_{e 4}}^2\abs{U_{\mu 4}}^2}{(1-\abs{U_{s4}}^2)^2} 
              \left( 1 + \frac{V_b - V_{\text{res}}}{A} \right) \sin^2\phi_2 \nonumber \\
    & &   + 2 \frac{\abs{U_{e 4}}^2\abs{U_{\mu 4}}^2}{(1-\abs{U_{s4}}^2)^2}
              \left( 1 - \frac{V_b - V_{\text{res}}}{A} \right) \sin^2\phi_3 \text{,} \label{eqn:P_mu_e} \\
 P_{\nu_{\mu} \rightarrow \nu_{b}}
    &=&  4 \abs{U_{\mu 4}}^2 \abs{U_{s4}}^2 
         \left(\frac{\Delta m_{41}^2}{2 E A}\right)^2 \sin^2\phi_1 \label{eqn:P_mu_s} \text{,} \\
 P_{\nu_{\mu} \rightarrow \nu_{\mu}}
    &=&  1 - P_{\nu_{\mu} \rightarrow \nu_{e}} - P_{\nu_{\mu} \rightarrow \nu_{b}} \text{,} \label{eqn:P_mu_mu}
\end{eqnarray}
where the oscillation phases are 
\begin{alignat}{2}
  \phi_1 &= \frac{\lambda_4-\lambda_3}{2}L &&= \frac{L}{2}A \text{,} \label{eqn:phi_1} \\
  \phi_2 &= \frac{\lambda_3}{2}L &&= \frac{L}{4} \left(V_b+\frac{\Delta m^2_{41}}{2E} -A\right) \text{,} \label{eqn:phi_2} \\
  \phi_3 &= \frac{\lambda_4}{2}L &&= \frac{L}{4}\left(V_b+\frac{\Delta m^2_{41}}{2E} +A\right) \label{eqn:phi_3}
\end{alignat}
and $V_{\text{res}}$ is the value of the matter potential at which $A$ takes its minimal
value $\abs{U_{s4}} \sqrt{1 - \abs{U_{s4}}^2} \, \Delta m_{41}^2 / E$. It is given by 
\begin{equation}
  V_{\text{res}} = -\frac{\Delta m_{41}^2}{2 E} \left(2 \abs{U_{s4}}^2 - 1\right) 
  \label{eqn:V_res_new}
\end{equation}
and corresponds to the new MSW resonance condition.  Whether the resonance is
in the neutrino or anti-neutrino sector depends on the sign of $V_b$, i.e.\ the
relative sign of the charges $g_b$ and $g_b'$. With the assumption $\sin^2
\theta_{24}<0.5$ and for $V_b<0$ ($V_b>0$) the resonance condition can be
fulfilled only in the neutrino (anti-neutrino) sector.  For $\Delta m_{41}^2 =
1$~eV$^2$, a matter density of 3~g/cm$^3$ and a neutrino energy of 1~GeV, the
resonance condition is fulfilled for neutrinos if $\epsilon = G_B / \sqrt{2} G_F
\simeq -2 \times 10^3$ and for anti-neutrinos if $\epsilon$ has opposite sign.
For oscillation experiments, we see that matter enhancement of active-to-sterile
neutrino oscillations is expected predominantly in high energy ($\mathcal{O}(\text{GeV})$)
experiments and only if the new gauge force is several orders of magnitude stronger
than SM weak interactions. For weaker gauge forces, the new resonance moves to higher
energies that are only accessible with atmospheric or cosmic neutrinos.

Note that equation~\eqref{eqn:V_res_new} has a structure similar to the
expression for the standard MSW resonance condition. To see this, consider the
matrix element $\abs{U_{s4}}^2$ in the parametrization of
equation~\eqref{eqn:parametrization}: $\abs{U_{s4}}^2 = \cos^2 \theta_{14}
\cos^2 \theta_{24} \cos^2 \theta_{34}$. If $\cos^2 \theta_{34}$, $\cos^2
\theta_{14} \approx 1$, we have $V_{\text{res}} = - (\Delta m_{41}^2 / 2 E)
\cos 2 \theta_{24}$.  
However, unless $\Delta m^2_{41}/2 E$ is much larger than $V_b$, oscillations at
short baseline cannot be approximately described in an effective two-flavour framework, unlike
the 3+1 model without non-standard matter effects. The reason is that, without the extra
matter term, three eigenvalues of the Hamiltonian can be set
to zero at short baseline, while large $V_b$ implies that this is only possible
for two of them.

On the other hand, in the limit of very large matter potential, $V_b \gg \Delta m_{41}^2/(2E)$, the
term proportional to $\sin^2 \phi_2$ in equation~\eqref{eqn:P_mu_e}
dominates over the terms containing $\sin^2 \phi_1$
and $\sin^2 \phi_3$ since the latter two are of higher order in $\Delta m^2_{41}/(2EV_b)$. If we
furthermore assume the baseline is not too long, in particular $(\Delta
m^2_{41})^2/(4 E^2 V_b) \cdot L/2 \ll 1$, we can approximate $\phi_2 \approx
(L/2) (1-\abs{U_{s4}}^2) \Delta m^2_{41}/(2E)$ and obtain for the oscillation
probability of equation~\eqref{eqn:P_mu_e} the effective two-flavour formula
\begin{eqnarray}
  P_{\nu_{\mu} \rightarrow \nu_{e}}
    &\approx& 4 \frac{\abs{U_{e 4}}^2\abs{U_{\mu 4}}^2}{ \left(1-\abs{U_{s4}}^2 \right)^2} \cdot
     \sin^2 \left(\frac{L}{2} \left( 1-\abs{U_{s4}}^2 \right) \frac{\Delta m_{41}^2}{2 E} \right) 
     + \mathcal{O}\left(\left(\frac{\Delta m^2_{41}}{2EV_b}\right)^2\right) \text{.} \label{eqn:P_mu_e_approx}
\end{eqnarray}
As expected, in the limit of large matter potential $V_b$, the corresponding
neutrino $\nu_b$ decouples from flavour oscillations, $P_{\nu_{\mu} \rightarrow
\nu_{b}} \approx 0$ and the $\nu_{\mu}$ survival probability becomes
$P_{\nu_{\mu} \rightarrow \nu_{\mu}} \approx 1 - P_{\nu_{\mu} \rightarrow
\nu_{e}}$.

We do not expect that scenarios with large $V_b$ can explain the short-baseline
anomalies better than conventional models without new interactions. The
reactor~\cite{Mueller:2011nm, Mention:2011rk, Huber:2011wv} and
gallium~\cite{Acero:2007su,Giunti:2010zu} experiments were too low in energy;
in LSND~\cite{Aguilar:2001ty}, neutrinos traveled mostly through air;
MiniBooNE could in principle be sensitive to new matter effects, but resonant
enhancement could only explain an anomaly in either the neutrino or the
anti-neutrino sector, while the data shows similar deviations from expectations
in both sectors.\footnote{Note that in earlier MiniBooNE
data~\cite{AguilarArevalo:2007it, AguilarArevalo:2009xn,
AguilarArevalo:2010wv}, there appeared to be mild tension between the neutrino
and anti-neutrino mode data. This motivated the authors
of~\cite{Karagiorgi:2012kw} to consider resonantly enhanced active--sterile
neutrino mixing even as a possible \emph{explanation} of the MiniBooNE
anomaly.} On the other hand, we expect that MiniBooNE---along with
long-baseline experiments like MINOS and with solar neutrinos---will impose
tight constraints on $V_b$.

\subsection{Accuracy of analytic approximations}
\label{subsec:ksc}

In the following, we discuss the implications of sterile neutrinos with
non-standard matter effects in terrestrial long-baseline experiments, taking
MiniBooNE and MINOS as examples.  In doing so, we also compare our analytic
expressions \eqref{eqn:P_mu_mu} and \eqref{eqn:P_mu_e} to a numerical
computation in the full four flavour framework and to the results of
\cite{Karagiorgi:2012kw}.

To obtain the exact four-flavour oscillation probabilities, we diagonalize the
effective Hamiltonian of equation~\eqref{eqn:effectiveH} numerically and use
the resulting eigenvalues and eigenvectors in equation~\eqref{eqn:OscProb}.  In
doing so, we absorb the neutral current potential $V_\text{NC}$ into a
redefinition of $V_b$.\footnote{This is only approximately correct if $V_b
\simeq V_\text{NC}$ and the proton-to-neutron ratio is varying along the
neutrino trajectory.  Since we are mainly interested in scenarios with $V_b \gg
V_\text{NC}$, our results are insensitive to this subtlety.}  To average out
fast oscillations that would not be resolvable by experiments, we also
implement a low-pass filter by multiplying each term in the oscillation
probability equation~\eqref{eqn:OscProb} by a Gaussian
factor~\cite{GLoBES:Manual}. This yields:
\begin{eqnarray}
 P_{ \nu_{\alpha} \rightarrow \nu_{\beta}} 
      &=& \sum_{j,k} \tilde{U}_{\alpha j}^* \tilde{U}_{\beta j} \tilde{U}_{\alpha k} \tilde{U}_{\beta k}^* 
      \, \exp\big[ -i L (\lambda_j - \lambda_k) \big]
      \, \exp\Big[- L^2 (\lambda_j - \lambda_k)^2 \cdot \frac{\sigma_f (E)^2}{2 E^2} \Big] \text{,}
 \label{eqn:low_pass_filter}
\end{eqnarray}
where $\sigma_f(E)$ is the energy width of the filter, which is related to the
energy resolution of the experiment. This form for the low-pass filter can also
be motivated in a wave packet treatment, where the finite energy resolution of
the production and detection processes determines the width of the neutrino
wave packets (see \cite{Beuthe:2001rc} and references therein).  When comparing
analytical and numerical results, we also apply such a low-pass filter to the
analytic expressions \eqref{eqn:P_mu_e} and \eqref{eqn:P_mu_mu} by replacing
the oscillation terms $\sin^2 \phi_i$ according to
\begin{eqnarray}
 \sin^2 \phi_i \mapsto \frac{1}{2} \bigg(1 - \cos (2 \phi_i) 
        \cdot \exp\Big[- (2 \phi_i)^2 \, \frac{\sigma_f(E)^2}{2 E^2} \Big] \bigg) \text{.}
\end{eqnarray}
In the following, we choose $\sigma_f(E)=0.01$~GeV.

\begin{figure}
 \centering
 \includegraphics[width=0.55\textwidth]{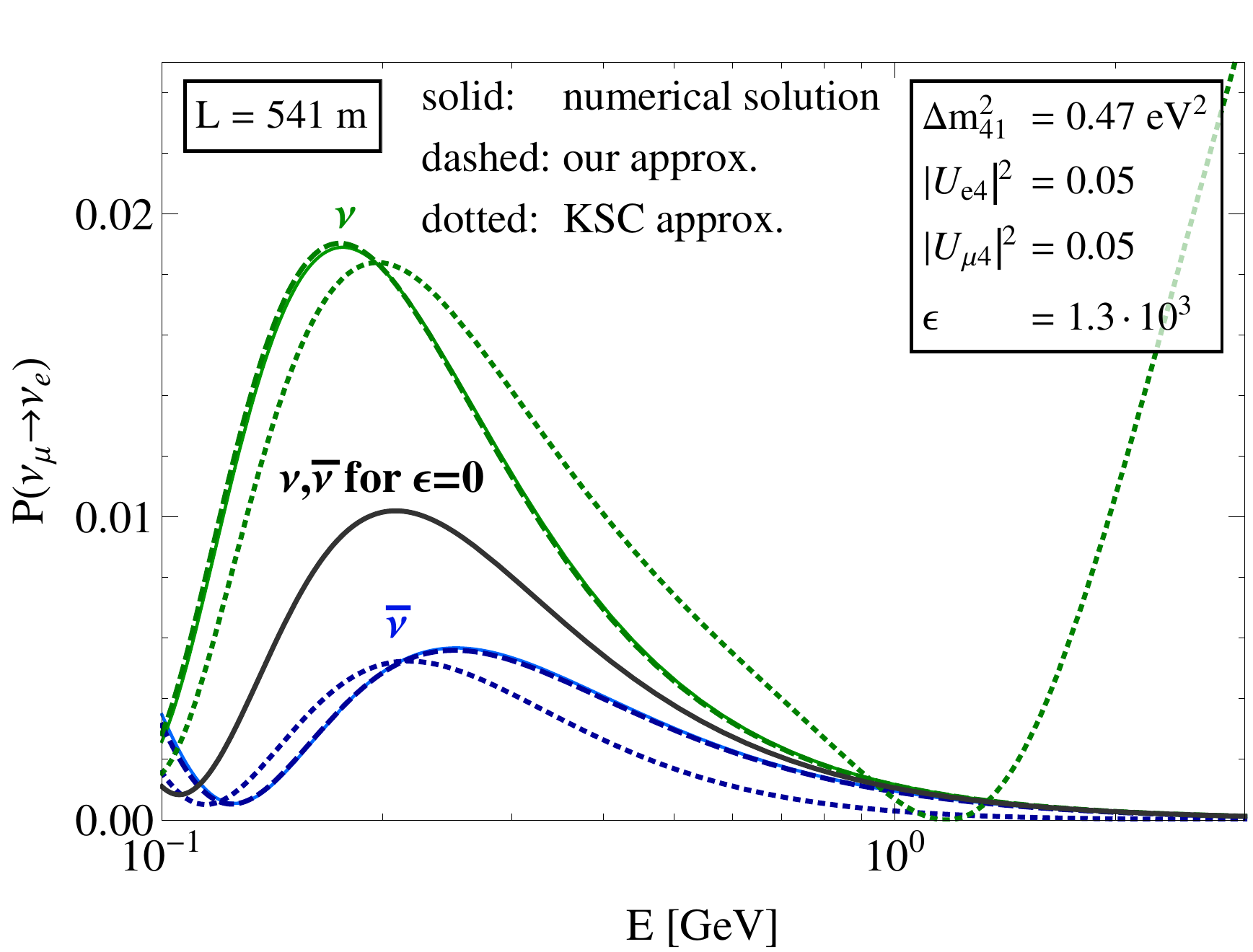}
 \caption{The electron neutrino (green) and anti-neutrino (blue) appearance
    probability in a model with a large MSW potential in the sterile sector (for
    instance the baryonic sterile neutrino model from~\cite{Pospelov:2011ha}).
    We use the baseline $L = 541$~m and the energy range 0.1--3~GeV of the
    MiniBooNE experiment and take the favored model parameters
    from~\cite{Karagiorgi:2012kw}: $\epsilon = G_B / (\sqrt{2} G_F) = 1.3 \cdot 10^3$
    ($\Leftrightarrow V_b = 2 \cdot 10^{-10} \, \text{eV}$ for $\rho=3$~g/cm$^3$),
    $\Delta m_{41}^2 = 0.47$~eV$^2$, $|U_{s4}|^2 = 0.9$, $|U_{e4}|^2 = |U_{\mu4}|^2 = 0.05$.
    (We will see below, that this particular parameter point is in fact excluded
    by MINOS data, though.) For the standard
    oscillation parameters, we have used the results of the global fit ``Free Fluxes and RSBL'' 
    of~\cite{GonzalezGarcia:2012sz}.  In black, we show also the prediction of a
    sterile neutrino model without new interactions ($\epsilon = 0$). Dashed lines
    correspond to our analytic approximations, which coincide with numerical
    results (solid curves) in this baseline and energy range, while dotted lines 
    show the results from \cite{Karagiorgi:2012kw}.}
 \label{fig:KSC_MiniBooNE_range_reproduced}
\end{figure}

In the calculation of the analytical formulas in~\cite{Karagiorgi:2012kw}
the eigenvalues $\lambda_i$ are approximated
by setting $A \approx V_b + \frac{\Delta m^2_{41}}{2E}$ (i.e.\ taking $|U_{s4}|
= 1$ in equation~\eqref{eqn:A}). This leads to $\lambda_1 = \lambda_2 =
\lambda_3=0$ and $\lambda_4=V_b + \Delta m^2_{41}/ 2E$.  The oscillation phases
of equations~\eqref{eqn:phi_1}--\eqref{eqn:phi_3} then become $\phi_1 = \phi_3 =
\frac{1}{2} L (V_b + \Delta m_{41}^2 / 2 E)$ and $\phi_2 = 0$. With this replacements 
our equation~\eqref{eqn:P_mu_e} reduces to equations~(21)--(22) in~\cite{Karagiorgi:2012kw}. 
In the limit of large $V_b$ we see from equation~\eqref{eqn:P_mu_e_approx} that this 
approximation is only valid if $L/2 ( 1-\abs{U_{s4}}^2) \Delta m_{41}^2/(2 E) \ll 1$.

Since the latter condition is fulfilled in the $L/E$ regime at which the LSND
and MiniBooNE experiments are sensitive to $\nu_{\mu} \rightarrow \nu_e$
flavour transitions, the approximation from \cite{Karagiorgi:2012kw} is
applicable there.  This can be seen in
figure~\ref{fig:KSC_MiniBooNE_range_reproduced}, where the transition
probabilities for neutrinos (in green) and anti-neutrinos (in blue) are shown for
$L = 541$~m and $E = 0.1$--3~MeV.  We have taken the model parameters at the
best fit point from~\cite{Karagiorgi:2012kw} (which we will show to be in fact
excluded by MINOS in section~\ref{sec:results}).  Dashed
curves correspond to our analytical approximation
(equation~\eqref{eqn:P_mu_e}), which agrees extremely well with numerical
results, while dotted curves show the approximation from equations~(21)--(22)
of~\cite{Karagiorgi:2012kw}.  The difference between the neutrino and
anti-neutrino sectors originates from the different signs of the matter
potential. As expected, $\epsilon > 0$ ($\Leftrightarrow V_b>0$) leads to a resonant enhancement of the
anti-neutrino transition probabilities and a suppression of the neutrino transition
probabilities compared to the case $\epsilon = 0$ (black curve).  We see that the
approximations used in \cite{Karagiorgi:2012kw} are fairly accurate in the most
relevant energy range below 1~GeV, but fail at higher energies.

\begin{figure}
  \centering 
  \includegraphics[width=\textwidth]{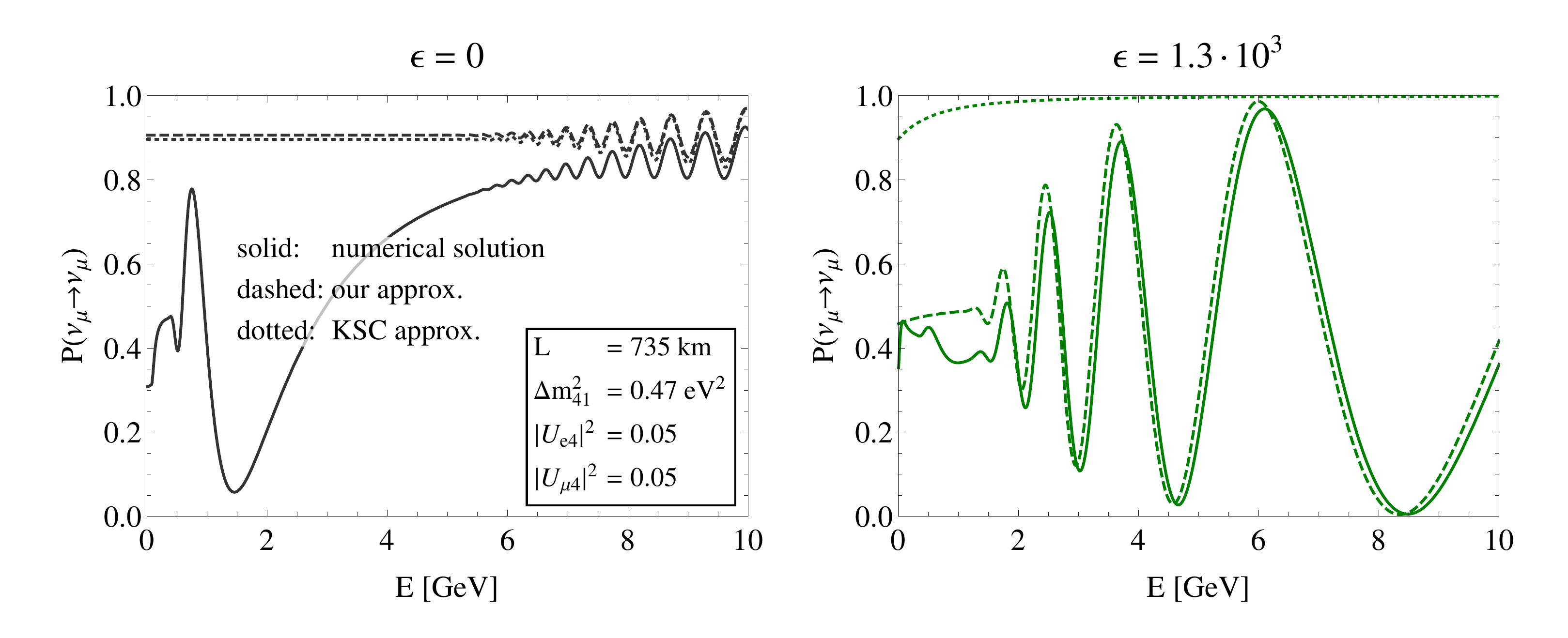}
  \caption{The figure shows the $\nu_{\mu}$ disappearance probability for the MINOS baseline 
           of $L=735 \, \text{km}$ and energies up to $10 \, \text{GeV}$.  We
           show the survival probability $P_{\nu_{\mu} \rightarrow \nu_{\mu}}$
           for $\epsilon=0$ (left panel) and $\epsilon = 1.3 \cdot 10^3$ 
           ($\Leftrightarrow V_b = 2 \cdot 10^{-10} \, \text{eV}$ for $\rho=3$~g/cm$^3$)
           (right panel) using the best fit parameters from the LSND/MiniBooNE
           fit of~\cite{Karagiorgi:2012kw}.  Solid curves correspond to a
           numerical calculation in the full four flavour oscillation
           framework, using for the standard oscillation parameters the values
           from the fit ``Free Fluxes and RSBL'' of~\cite{GonzalezGarcia:2012sz}.
           Dashed curves show our
           analytic approximation, equation~\eqref{eqn:P_mu_mu}, while dotted
           curves correspond to equations 20 and 28
           of~\cite{Karagiorgi:2012kw}.  The comparison shows that, when the
           new matter potential $V_b$ is switched on ($\epsilon>0$), the active-sterile
           oscillation mode dominates over the standard atmospheric oscillation
           pattern, an effect which is not captured by the approximations
           made in \cite{Karagiorgi:2012kw}.}
  \label{fig:KSC_vs_My_Approx}
\end{figure}

Since standard and non-standard matter effects are most relevant at long
baseline ($\gtrsim \text{few} \times 100$~km), it is important to also study
the disappearance probability $1 - P_{\nu_{\mu} \rightarrow \nu_{\mu}}$
as a function of energy for long-baseline oscillation
experiments like MINOS. MINOS has measured $P_{\nu_{\mu} \rightarrow
\nu_{\mu}}$ at a baseline of $L=735$~km in the energy range 1--50~GeV.
The oscillation probabilities for this baseline and energy range are shown
in figure~\ref{fig:KSC_vs_My_Approx} for $\epsilon > 0$ (right panel) and also for
the Standard Model ($\epsilon = 0$, left panel). We see that, due to matter-enhanced oscillations
inside the earth, a scenario with strong
non-standard matter effects leads to very large muon disappearance even at
energies as high as 10~GeV, well above the standard oscillation maximum at $\sim 1.5$~GeV.
This is in conflict with MINOS data and we therefore
expect that MINOS is able to place very strong constraints on new matter effects
in the sterile neutrino sector.  Figure~\ref{fig:KSC_vs_My_Approx} also implies
that the parameters favored in \cite{Karagiorgi:2012kw} are ruled out by MINOS.

Comparing numerical results (solid lines) to our analytic approximation (dashed lines), we
find, as expected, that the approximations of equations~\eqref{eqn:P_mu_e} and
\eqref{eqn:P_mu_mu} are accurate at $\Delta m_{31}^2 L / (2 E) \sim 1$ only
if $V_b$ is very large.  We also see that the analytic approximations from
\cite{Karagiorgi:2012kw} (dotted curves in figure~\ref{fig:KSC_vs_My_Approx})
are not applicable at long baseline even for large $V_b$. For example, in the
MINOS case, $\Delta m_{41}^2 L / (2 E) \cdot \,
\text{eV}^2 \sim 100$ for $\Delta m_{41}^2 \sim \text{eV}^2$,
the phase $\phi_2$ (see equations~\eqref{eqn:phi_2} and~\eqref{eqn:P_mu_e_approx})
becomes non-negligible. This is the reason
why our conclusions regarding the importance of MINOS data for
constraining sterile neutrino matter effects differ from those of
\cite{Karagiorgi:2012kw}, where $\phi_2$ has been neglected.

\section{Constraints from oscillation experiments}
\label{sec:osc-constraints}

From the analysis in the previous section we expect that the baryonic sterile
neutrino model (or models with new sterile neutrino--SM interactions in
general) could potentially explain by resonant enhancement an event excess in
the MiniBooNE neutrino or anti-neutrino data (but not in both), but is strongly
constrained by data from long-baseline experiments.
Therefore, we now derive limits on the model using a numerical $\chi^2$
analysis of data from MiniBooNE, MINOS and also solar neutrino experiments.

\subsection{Analysis method}
\label{sec:analysis-method}

In our analysis we fix the standard oscillation parameters at their best fit
values from the global fit by Gonzalez-Garcia et
al.~\cite{GonzalezGarcia:2012sz} (see table~\ref{tab:Parameters}) and we assume
a normal mass ordering. We have checked that our results for inverted ordering
are very similar, with only the solar limits becoming somewhat weaker. (We will comment
on this in more detail in section~\ref{sec:results}.)
For simplicity we set $\delta_{13}=\delta_{14}=\delta_{24}=0$ because none of
the experiments considered here is sensitive to CP violation in the small $V_b$
limit and equations~\eqref{eqn:P_mu_e}--\eqref{eqn:P_mu_mu} show that also the
leading terms in the oscillation probabilities for large $V_b$ are independent
of complex phases.   We fix the mixing angle $\theta_{34}=0$ since MiniBooNE is
not sensitive to this angle and MINOS has only very limited
sensitivity~\cite{Kopp:2013vaa}.  The impact of $\theta_{34}>0$ on the
constraints from solar experiments will be discussed in section~\ref{sec:results}.
Finally, we set $\sin^2 2\theta_{14}=0.12$ so that the reactor
anomaly~\cite{Mueller:2011nm,Mention:2011rk,Huber:2011wv} can be explained.
We will comment on the effect of relaxing this assumption also in section~\ref{sec:results}.
The constraints we impose on the parameter space are also summarized in
table~\ref{tab:Parameters}.  The remaining three parameters $\epsilon = G_B/(\sqrt{2} G_F)$, $\Delta
m^2_{41}$ and $\theta_{24}$ are scanned over the ranges $\abs{\epsilon}=1 -
32000$, $\Delta m^2_{41}/eV^2= 0.01 - 11$ and $\sin^2 \theta_{24}=0.0001 - 1$.

\begin{table}
  \centering
  \begin{ruledtabular}
  \begin{tabular}{cccccccc}
    $\sin^2 \theta_{12}$  &  $\sin^2 \theta_{23}$  &  $\sin^2 \theta_{13}$  &
    $\Delta m_{21}^2 \, [\text{eV}^2]$  &  $\Delta m_{31}^2 \, [\text{eV}^2]$  &
    $\delta_{13}$, $\delta_{14}$ $\delta_{24}$  &  $\sin^2 2 \theta_{14}$  & 
    $\sin^2 \theta_{34}$ \\ \hline
    $0.302$  &  $0.413$  &  $0.0227$  &  $7.5 \cdot 10^{-5}$  &  $2.473 \cdot 10^{-3}$ &
    $0$  &  $0.12$  &  $0$
  \end{tabular}
\end{ruledtabular}
  \caption{The parameter values of the baryonic sterile neutrino model that we
    have fixed in our parameter scan.} 
  \label{tab:Parameters}
\end{table}

We now discuss the details of our fits to MINOS, MiniBooNE and solar neutrino data.

\subsubsection{MINOS}

For MINOS, we use GLoBES~\cite{Huber:2004ka, Huber:2007ji} to compute the
energy dependent oscillation probabilities $P_\text{near}(E)$ for the near
detector and $P_\text{far}(E)$ for the far detector numerically. We include a
low pass filter according to equation~\eqref{eqn:low_pass_filter} with
$\sigma_f(E)=0.06 \cdot E$.  The matter density $\rho$ along the neutrino
trajectory to the far detector is assumed to be constant at its average value
\begin{eqnarray}
  \ev{\rho_\text{far}} &=& \frac{2}{L_\text{far}} \int_{\sqrt{R_\oplus^2-(L_{\text{far}}/2)^2}}^{R_\oplus}
  \rho (r) \frac{d}{dr} \Big( \sqrt{r^2-R_\oplus^2+(L_{\text{far}}/2)^2} \Big) dr \text{.}
\end{eqnarray}
In this expression, which can be understood from geometric arguments, $r$ is
the distance of the neutrino from the center of the earth, $R_\oplus$ is the radius of
the earth and $L_{\text{far}}=735 \, \text{km}$ is the neutrino path length
from the source to the far detector~\cite{Michael:2006rx}. Using the matter
density profile from the Preliminary Reference Earth Model
(PREM)~\cite{Anderson1989} we obtain $\ev{\rho_{\text{far}}} \approx 2.36 \,
\text{g}/\text{cm}^3$. 

For large $V_b$, matter effects can be relevant even in the near detector at a
baseline $L_{\text{target}}=965 \, \text{m}$ from the target.  In computing the
average matter density $\ev{\rho}_{\text{near}}$ that neutrinos experience on
their way to the near detector, we account for the fact that they first travel
along the evacuated decay pipe with a length of $L_{\text{pipe}}=675 \,
\text{m}$.  We estimate $\ev{\rho}_{\text{near}} \approx (L_{\text{target}} -
L_{\text{pipe}}) / L_{\text{near}} \cdot 3\frac{\text{g}}{\text{cm}^3}$, where
$L_{\text{near}} \simeq 763 \, \text{m}$ is the average distance between the
neutrino production vertex and the near detector. It is obtained from
the decay length of the neutrinos' parent pions, which have an
average energy of $4 - 5 \, \text{GeV}$~\cite{Diwan:2004cb}.

We compute the theoretically predicted event spectrum $N_{\text{osc}}$  by 
multiplying the ratio $P_{\text{far}}(E) / P_{\text{near}}(E)$ 
with the background-subtracted prediction for the MINOS event rate in the absence
of oscillations, $N_{\text{no osc}}$:
\begin{eqnarray}
 N_{\text{osc}}(E) = \big[ N_{\text{no osc}}(E) - N_{\text{bg}}(E) \big]
   \frac{P_{\text{far}}(E)}{P_{\text{near}}(E)}\text{.}
 \label{eqn:N_th}
\end{eqnarray}
The no-oscllation rate $N_{\text{no osc}}(E)$ and the background rate $N_{\text{bg}}(E)$
are taken from~\cite{deJong:2013-04-19pya}, which is similar to~\cite{Adamson:2013whj} 
but contains data up to $50 \, \text{GeV}$. The higher energy data is important
to us since it increases the sensitivity at low matter potential $V_b$.

To account for the finite energy resolution of the detector, we fold $N_\text{osc}$
with the detector response function $f(E,E')$, which maps the
true event energy $E'$ to the reconstructed energy $E$.  Finally, we also
add the small experimental background $N_\text{bg}(E)$:
\begin{eqnarray}
  N_{\text{th}}(E) = N_{\text{bg}} (E) + \int \! f(E,E') N_{\text{osc}}(E') dE' \text{.}
 \label{eqn:N_rec}
\end{eqnarray}
We assume a Gaussian shape for $f(E,E')$,
\begin{eqnarray}
  f(E,E') = \frac{1}{\sigma(E')\sqrt{2\pi}} \exp\left(- \frac{(E-E')^2}{2 \sigma^2(E')}\right) \text{,}
  \label{eqn:smearing_prob}
\end{eqnarray}
where we choose $\sigma(E') = 0.2 \, \text{GeV} \sqrt{E'/\text{GeV}}$.  This
choice allows us to reproduce the oscillated event rates and the constraints on
$\theta_{23}$ and $\Delta m_{31}^2$  from \cite{Adamson:2013whj} with good
accuracy.  When evaluating equation~\eqref{eqn:N_rec} numerically, we
discretize the integral so that $N_\text{osc}(E)$ needs to be evaluated only at
fixed support points $E_j'$ with a step size of $\Delta E_j' = 0.25 \,
\text{GeV}$ in between.  (We have checked that choosing a smaller value for $\Delta E_j'$
does not change our results significantly, which implies that possible aliasing
effects are under control.) Following the MINOS
analysis~\cite{deJong:2013-04-19pya}, events are binned for the analysis
according to their reconstructed energy $E$. The rate in th $i$-th bin is given
by
\begin{align}
  N_\text{th}^i = \int_{E_i-\Delta E_i /2}^{E_i+\Delta E_i /2} N_\text{th}(E) \, dE
                = N_\text{bg}^i + \sum_j F^{ij} N_\text{osc}(E'_j) \text{,}
  \label{eqn:N_rec-i}
\end{align}
where $N_\text{bg}^i$ is the total background in the $i$-th bin and
the elements of the detector response matrix $F^{ij}$ are
$F^{ij} \equiv = \int_{E_i-\Delta E_i /2}^{E_i+\Delta E_i /2} f(E,E_j') \, dE$.
It is important to note that the $F^{ij}$ need to be computed only once.

From equation~\eqref{eqn:N_rec-i} we compute $\chi^2$ according to 
\begin{eqnarray}
 \chi^2 = \sum_i \Bigg(
 \frac{ N_\text{th}^i - N_\text{exp}^i }
      { \sqrt{N_\text{exp}^i} + 0.1 \cdot N_\text{exp}^i } \Bigg)^2 \text{,}
\end{eqnarray}
where $N_{\text{exp}}^i$ is the observed event rate in the $i$-th energy
bin~\cite{deJong:2013-04-19pya} and the sum runs over all energy bins.  Note
that we have included an additional uncertainty of $10\%$ in order to account
for systematic errors without modeling them in detail.  Like our choice of
$\sigma(E')$ in equation~\eqref{eqn:smearing_prob}, also our simplified
treatment of systematic errors has been confirmed by cross-checking our
simulations against the results of~\cite{Adamson:2013whj, Kopp:2013vaa}.

\begin{figure}
 \centering
 \includegraphics[width=\textwidth]{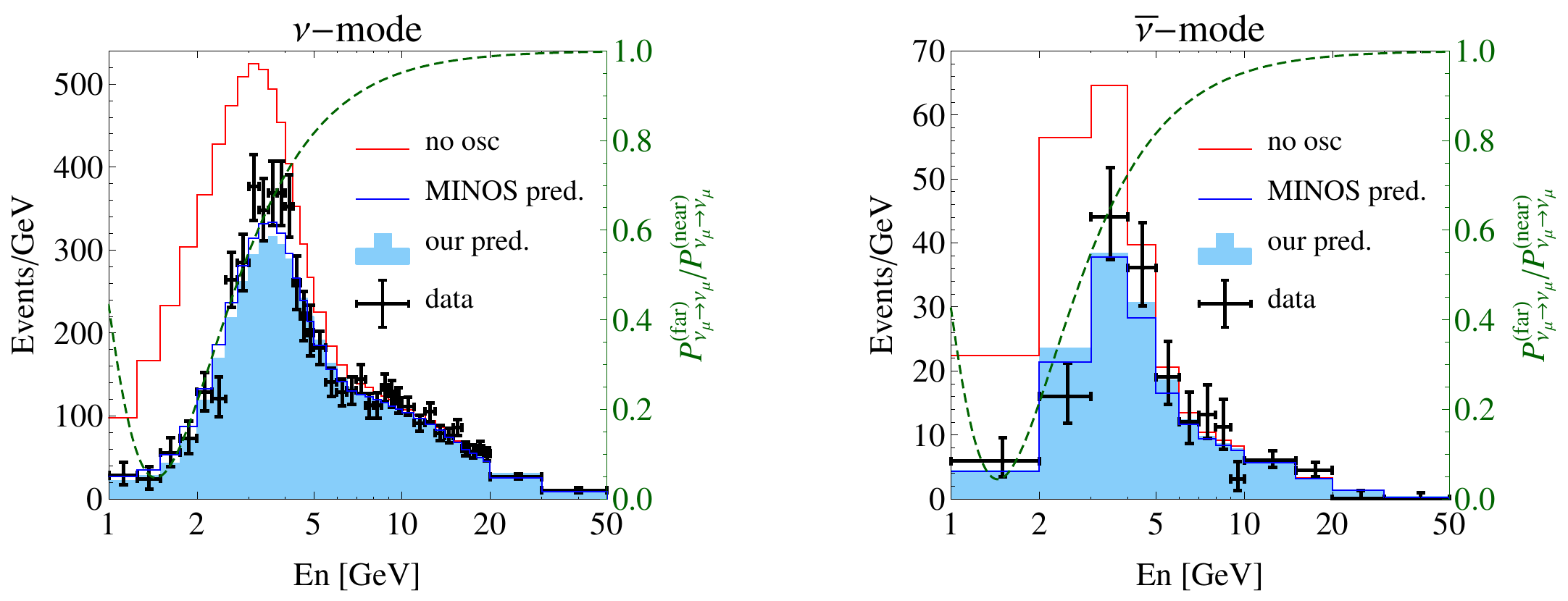}
 \caption{The measured and predicted event spectra for the MINOS ${\nu}_{\mu}$ (left) and 
          $\overline{\nu}_{\mu}$ (right) disappearance measurements. The red
          histogram is the MINOS prediction assuming no neutrino
          oscillation~\cite{deJong:2013-04-19pya}. In blue, we show the
          predicted event spectrum including oscillations according to
          equation~\eqref{eqn:N_rec}, assuming standard three flavour
          oscillations with the parameters listed in
          table~\ref{tab:Parameters}. The blue shaded histogram is our
          prediction, the unshaded histogram is the prediction by the MINOS
          collaboration.  We overlay the survival probability $P_{\nu_{\mu}
          \rightarrow \nu_{\mu}}$ (dashed green curve and vertical scale on
          the right).}
 \label{fig:MinosSpectrum}
\end{figure}

In figure~\ref{fig:MinosSpectrum}, we compare our prediction for the oscillated
neutrino spectrum in MINOS assuming standard 3-flavour oscillations (blue
shaded histogram) to the official MINOS prediction (blue unshaded histogram)
and to the data (black points with error bars). We find excellent agreement,
which validates our calculations. We also show the MINOS no oscillation
prediction (red histogram) which is the starting point for our predictions,
as well as the survival probability $P_{\nu_\mu \to \nu_\mu}$ (dashed green
line; corresponding vertical scale shown on the right).

\subsubsection{MiniBooNE}

As for MINOS, the oscillation probabilities for MiniBooNE are calculated
numerically in the full four flavour framework with the help of
GLoBES~\cite{Huber:2004ka, Huber:2007ji}, including a low pass filter according to
equation~\eqref{eqn:low_pass_filter} with $\sigma_f(E) = 0.06 E$.  Since the
MiniBooNE decay pipe is only 50~m long, while the distance
from the target to the detector is $L=541$~m, we neglect the effect
of the finite pion decay length. Instead, we take the matter density to
be $\ev{\rho} \sim 3 \, \text{g}/\text{cm}^3$ along the whole neutrino trajectory.

\begin{figure}
 \centering
 \includegraphics[width= \textwidth]{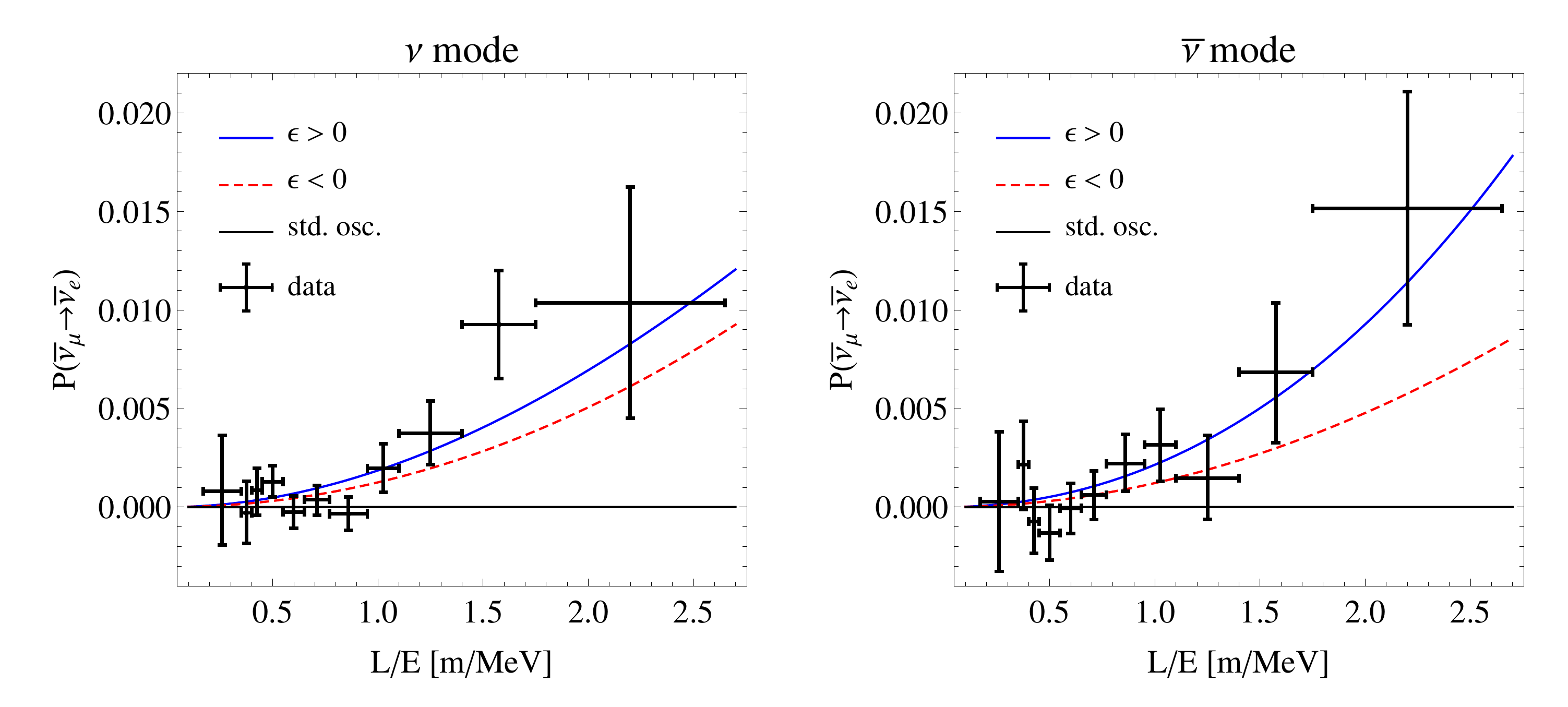}
 \caption{The measured MiniBooNE $\nu_\mu \to \nu_e$ (left) and $\bar\nu_\mu
          \to \bar{\nu}_e$ (right) appearance probabilities compared to the predictions
          of the baryonic sterile neutrino scenario for $\epsilon > 0$ (blue line) and
          $\epsilon < 0$ (dashed red line) at the MiniBooNE best fit points from table~\ref{tab:best_fit}.
          Without sterile neutrinos, the appearance probability at the MiniBooNE baseline
          is approximately zero (solid black line).}
 \label{fig:MinibooneSpectrum}
\end{figure}

We use a $\chi^2$ analysis to compare our predicted oscillation probabilities
with the experimentally measured probabilities, which are given
in~\cite{Aguilar-Arevalo:2013pmq} as a function of $L/E$.  The data from
\cite{Aguilar-Arevalo:2013pmq} are shown in figure~\ref{fig:MinibooneSpectrum}
together with the trivial no-oscillation prediction and with our prediction for
the MiniBooNE best fit points in the baryonic
sterile neutrino scenario for $\epsilon > 0$ and $\epsilon < 0$.

\subsubsection{Solar neutrinos}
\label{sec:solar}

We analyze solar neutrino oscillation data by comparing the
measured $\nu_e$ survival probability $P_{\nu_e \to \nu_e}$ at
different energies to our theoretical predictions.  The data points are taken
from~\cite{Bellini:2013lnn} and include results from
Super-Kamiokande, SNO, Borexino and radiochemical experiments.

In calculating $P_{\nu_e \to \nu_e}$, we assume MSW flavour transitions to be
fully adiabatic and we account for the fact that solar
neutrinos arrive at the earth as an incoherent mixture of mass eigenstates.  We
obtain $P_{\nu_e \to \nu_e}$ according to
\begin{eqnarray}
  \label{eqn:SolarAdiabFormula}
  P_{ \nu_{e} \rightarrow \nu_{e}} 
    = \sum_{i} \lvert U_{e i} \rvert^2 \cdot \lvert \tilde{U}_{e i}(0) \rvert^2 \text{,}
\end{eqnarray}
where $\tilde{U}_{e i}(0)$ is the mixing matrix in matter at the center of the
Sun ($t=0$) and $U_{e i}$ is the vacuum mixing matrix. We neglect earth matter
effects here, but we have checked that, in the parameter ranges of interest to
us, the day--night effect caused by the earth matter is of the order of few per
cent, comparable to the day--night effect in the Standard Model.  We thus
anticipate that our limits would only change marginally if Earth matter
effects were included.

In order to verify that the assumption of full adiabaticity is justified, we
have examined the adiabaticity parameter $\gamma$ in the two flavour
approximation and we have checked that the adiabaticity
condition~\cite{Akhmedov:1999uz}
\begin{eqnarray}
  \gamma^{-1} = \frac{\sin 2 \theta \frac{\Delta m_{ij}^2}{2E}}{\abs{\lambda_i-\lambda_j}^3} 
           \cdot \bigg| \frac{dV_{b}}{dt} \bigg|  \ll 1
\end{eqnarray}
holds for all relevant mass squared difference $\Delta m_{ij}^2$ even for large $V_b$ and the smallest relevant differences between the
Hamiltonian eigenvalues $\lambda_i$ and $\lambda_j$, which occur at the resonance position.
We determine the derivative of the matter potential, $\big| dV_{b}/dt \big|$, from the solar
density profile of the standard solar model BS'05
(OP)~\cite{Bahcall:2004pz}. 

\begin{figure}
 \centering
 \includegraphics[width=0.45\textwidth]{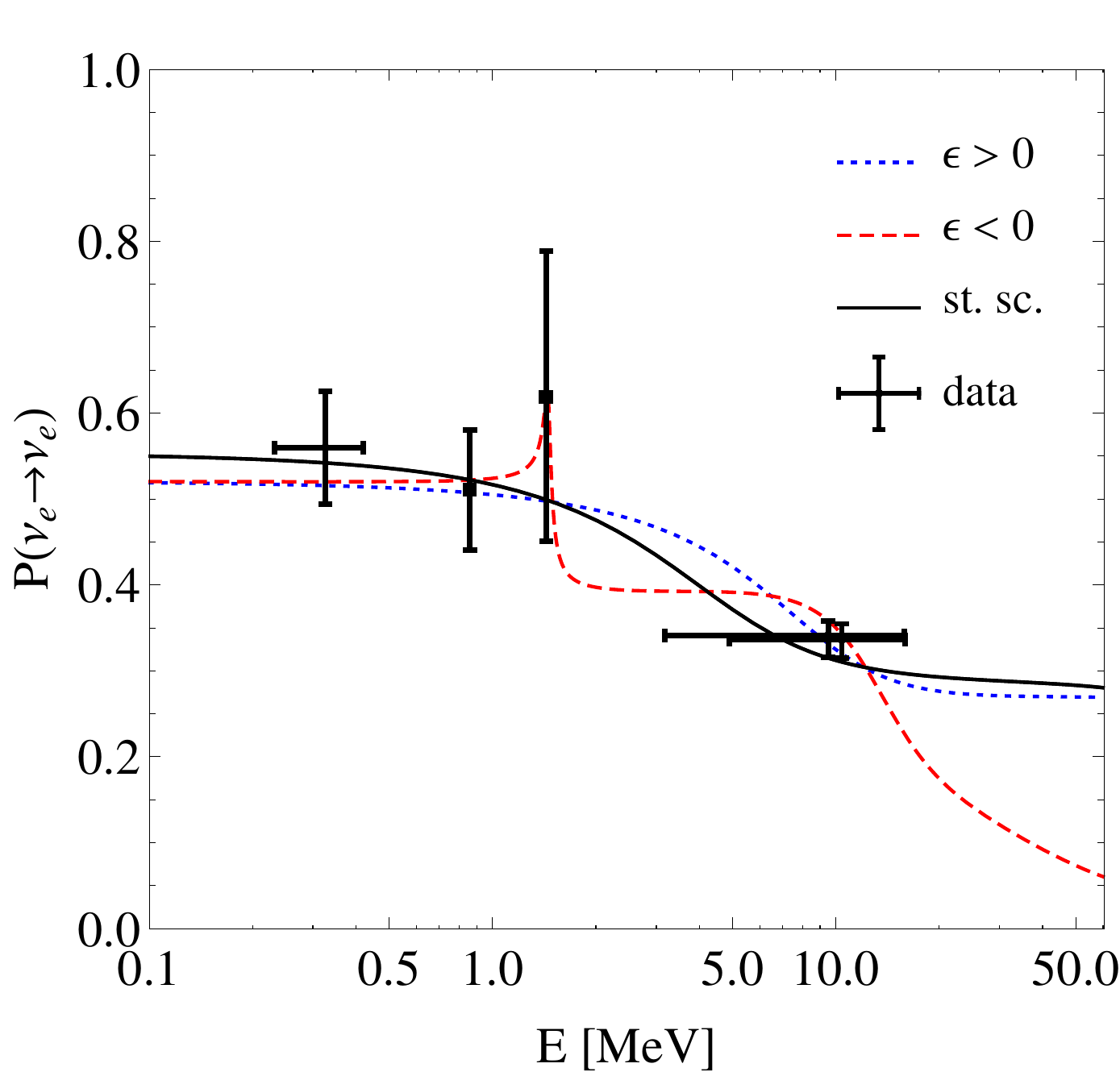}
 \caption{Comparison of the measured solar neutrino oscillation probabilities to our theoretical
          predictions for standard three flavour oscillations (black) and for the
          best fit parameter points of the baryonic sterile neutrino model with
          $\epsilon > 0$ (dotted blue) and $\epsilon < 0$ (dashed red).}
 \label{fig:SolarSpectrum}
\end{figure}

In figure~\ref{fig:SolarSpectrum}, we compare the measured solar neutrino oscillation
probabilities $P_{\nu_e \to \nu_e}$ to the theoretical predictions for standard three flavour oscillations
and for the best fitting baryonic neutrino scenarios with $\epsilon > 0$ (blue) and
$\epsilon < 0$ (red).

We observe that for $\epsilon < 0$, a peak-like structure appears in
$P_{\nu_e \to \nu_e}$, which suggests that mixing of $\nu_e$ with other flavors is
dynamically driven to zero for specific parameter combinations.  The peak occurs
at parameter points where $\Delta m_{41}^2 / (2 E),\ \Delta m_{31}^2 / (2 E) \gg
\Delta m_{21}^2 / (2 E)$, $V_b$ and where moreover $\theta_{34}$ and $\theta_{13}$ are small.
To understand its origin, it is therefore
helpful to determine the eigenvalues of the Hamiltonian $H_\text{eff}^\text{flavor}$
(see equation~\eqref{eqn:effectiveH}) using time-independent perturbation
theory, with the zeroth order Hamiltonian given by
\begin{align}
  H_\text{eff}^\text{flavour,(0)} \equiv 
    \frac{1}{2 E} U \diag(0, 0, \Delta m_{31}^2, \Delta m_{41}^2) U^\dag \,,
\end{align}
and the perturbation being $H_\text{eff}^\text{flavour,(1)} \equiv
H_\text{eff}^\text{flavour} - H_\text{eff}^\text{flavour,(0)}$. In the approximation
$\theta_{34} = \theta_{13} = 0$, a set of zeroth order
eigenvectors is obviously given by the matrix $U^{(0)} \equiv R_{24}' R_{14}' R_{23}$,
where, as before, $R_{ij}$ and $R_{ij}'$ are real and complex rotation matrices,
respectively.  Since $H_\text{eff}^\text{flavour,(0)}$ has zero as a double eigenvalue,
we next have to find eigenvectors of $H_\text{eff}^\text{flavour,(1)}$ in the
subspace corresponding to this double eigenvalue.
In other words, we need to compute $U^{(0) \dag} H_\text{eff}^\text{flavour,(1)}
U^{(0)}$ and then diagonalize the upper left $2 \times 2$ block.
It turns out that, if the condition
\begin{align}
  \frac{\Delta m_{21}^2}{2 E} \sin 2\theta_{12} + 
    V_b \cos\theta_{23} \sin\theta_{14} \sin 2\theta_{24} \simeq 0
  \label{eq:solar-peak-condition}
\end{align}
is fulfilled, this $2 \times 2$ block is automatically diagonal.  This implies that
$U^{(0)} (1, 0, 0, 0)^T \simeq (1, 0, 0, 0)$ is an approximate eigenvector of
$H_\text{eff}^\text{flavour}$.  Hence, if \eqref{eq:solar-peak-condition} holds at the
center of the Sun, solar neutrinos are produced in an almost pure $\nu_1$ mass eigenstate.
After adiabatic flavour conversion, the resulting $\nu_e$ admixture is of order
$\cos\theta_{12}^2$, leading to a peak in the observed solar neutrino spectrum
at Earth.

\subsection{Results} \label{sec:results}

\begin{figure}
  \centering
  \includegraphics[width=0.9\textwidth]{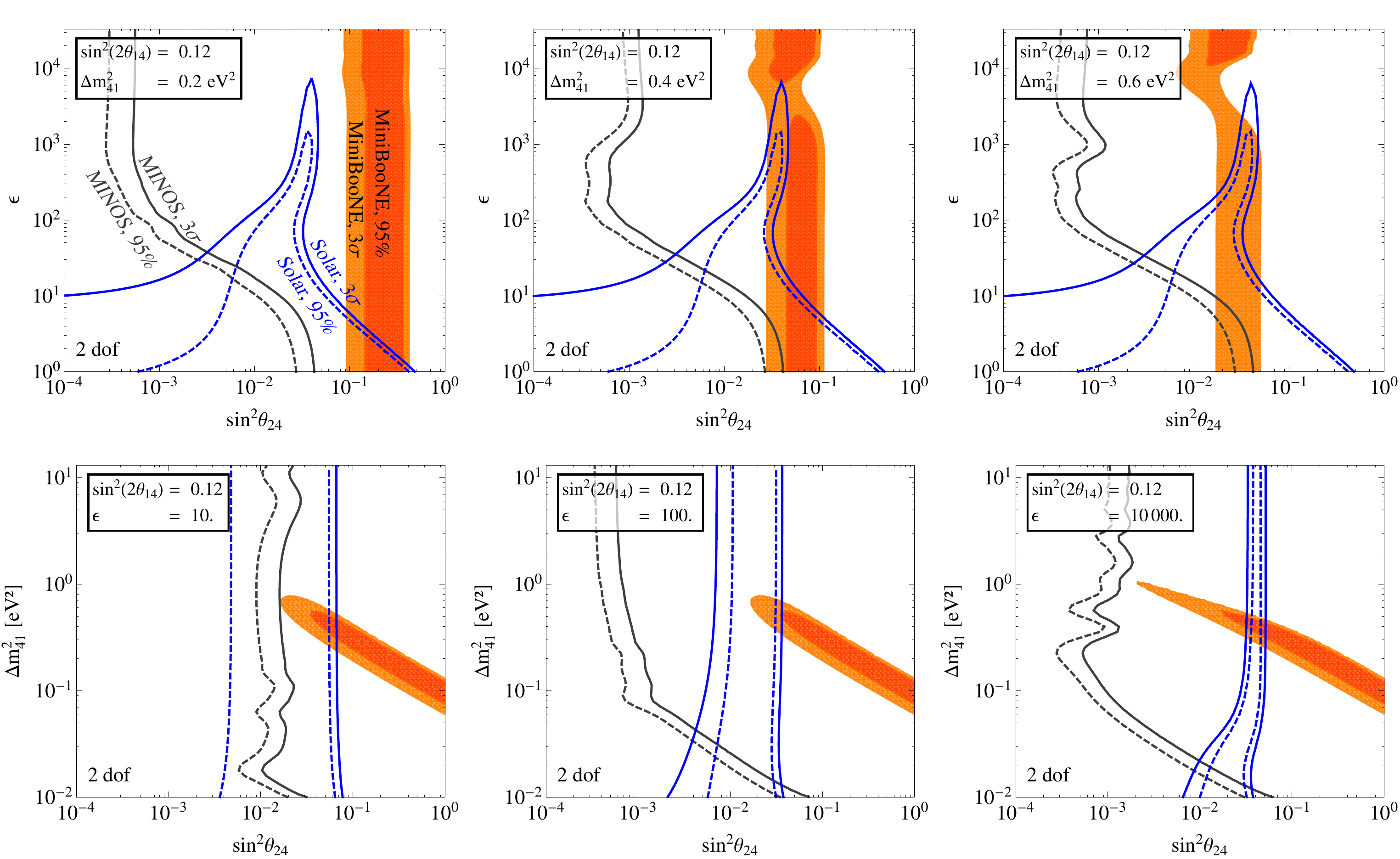}
  \caption{95\% and $3\sigma$ confidence level constraints on the parameters $\Delta m_{41}^2$,
    $\sin^2 2\theta_{24}$ and $\epsilon$ (strength of the new MSW potential) of the
    baryonic sterile neutrino model in the $\epsilon > 0$ case.  Blue contours show constraints
    from solar experiments, black contours are for MINOS and shaded areas correspond
    to the region preferred by MiniBooNE.
    We have fixed $\sin^2 2\theta_{14} = 0.12$, as motivated by the reactor and gallium
    anomalies.}
  \label{fig:ResultsEPSpos}
\end{figure}

\begin{figure}
  \centering
  \includegraphics[width=0.9\textwidth]{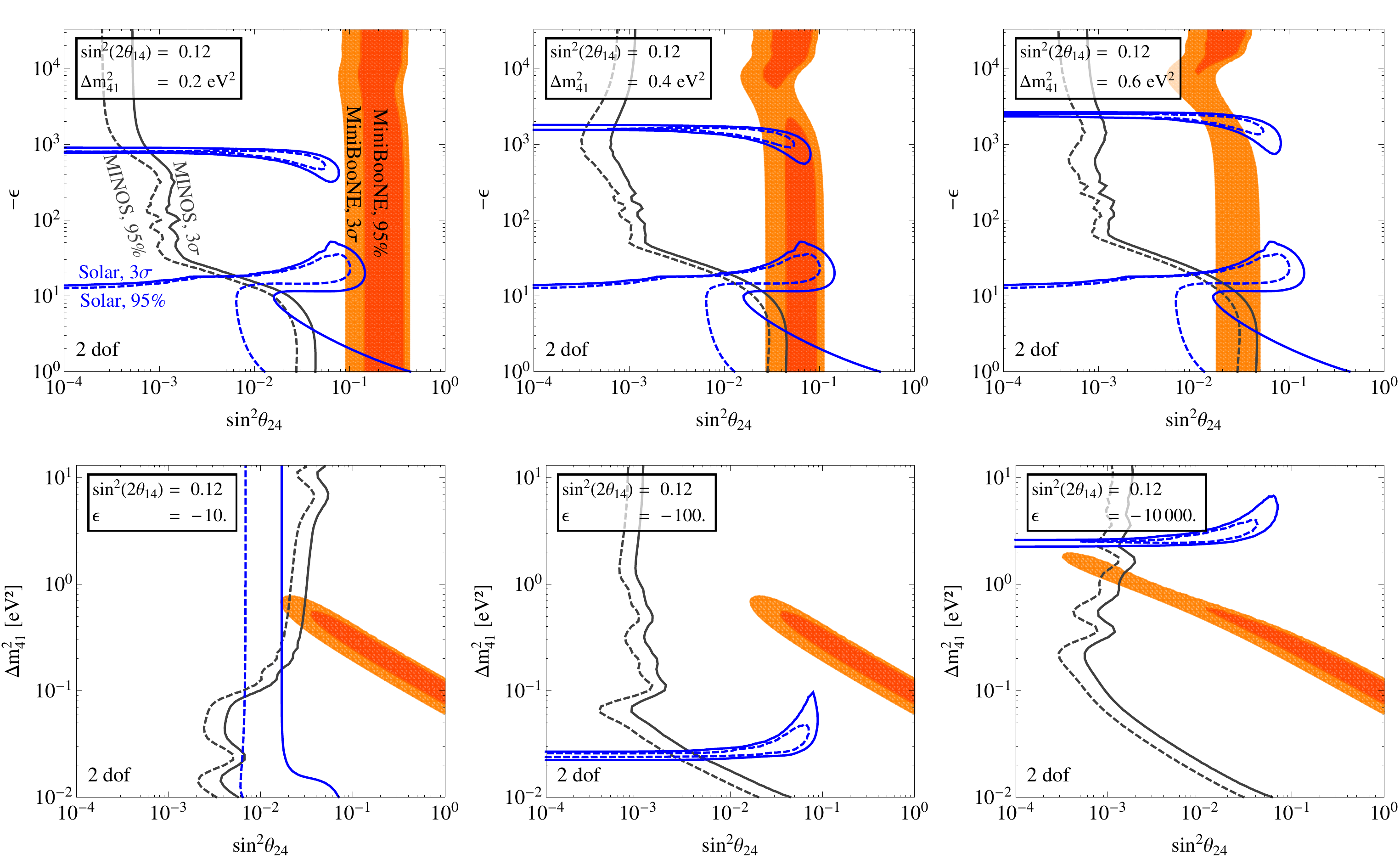}
  \caption{95\% and $3\sigma$ confidence level constraints on the parameters $\Delta m_{41}^2$,
    $\sin^2 2\theta_{24}$ and $\epsilon$ (strength of the new MSW potential) of the
    baryonic sterile neutrino model in the $\epsilon < 0$ case.  Blue contours show constraints
    from solar experiments, black contours are for MINOS and shaded areas correspond
    to the region preferred by MiniBooNE.
    We have fixed $\sin^2 2\theta_{14} = 0.12$, as motivated by the reactor and gallium
    anomalies.}
  \label{fig:ResultsEPSneg}
\end{figure}

In figures~\ref{fig:ResultsEPSpos} and \ref{fig:ResultsEPSneg} our constraints
on the parameter space of baryonic sterile neutrinos are presented as contour
plots for $\epsilon > 0$ and $\epsilon < 0$, respectively.  We show exclusion limits
(lines of constant $\chi^2 - \chi^2_{\text{min}}$) at the 95\% and $3\sigma$
confidence levels. In each panel, we keep either $\epsilon$ or $\Delta
m_{41}^2$ fixed at the value indicated in the plot 
and show constraints on the remaining two parameters.  Moreover,
as discussed in section~\ref{sec:analysis-method}, we fixed $\sin^2 2\theta_{14} = 0.12$.
Blue lines correspond to constraints from solar experiments, black lines are the
limits from MINOS and the colored regions show the parameter region preferred
by MiniBooNE.  The best fit values for $\epsilon>0$ and $\epsilon<0$ are listed in
table~\ref{tab:best_fit}.

\begin{table}
  \centering
  \begin{ruledtabular}
  \begin{tabular}{llllll}
             &              & $\epsilon=G_B/(\sqrt{2}G_F)$  & $\Delta m_{41}^2 \, [\text{eV}^2]$ & $\sin^2 \theta_{24}$& $\chi^2_{\text{min}} / \text{d.o.f.}$  \\ \hline
   MINOS     & $\epsilon>0$ & $ 16.9 $                      & $0.014$                            & $0.0024$            & $37.7 / 49$  \\
             & $\epsilon<0$ & $ -19.2 $                     & $0.037$                            & $0.00083$           & $36.1 / 49$  \\ \hline
   MiniBooNE & $\epsilon>0$ & $ 30634 $                     & $0.316$                            & $0.10$              & $16.1 / 20$  \\
             & $\epsilon<0$ & $ -32000 $                    & $0.116$                            & $0.75$              & $16.4 / 20$  \\ \hline
   Solar     & $\epsilon>0$ & $ 0.20 $                      & insensitive                        & $1.0$              & $1.10 / 3$   \\
             & $\epsilon<0$ & $ -38.0 $                     & $0.013$                            & $0.046$             & $0.41 / 3$        
  \end{tabular}
  \end{ruledtabular}
  \caption{Best fit values resulting from our parameter scan
    for the different experimental data sets. For the MiniBooNE fit with
    $\epsilon<0$ analysis the best fit value for $\epsilon$ is located outside the boundary 
    of the analysis region, but $\chi^2$ hardly depends on $\abs{\epsilon}$ in this region.
    Also note that the solar best fit in the $\epsilon>0$ case has $\epsilon<1$ 
    and is not sensitive to the exact value of $\Delta m_{41}^2$ in the interval $[0.01,11]$.}
  \label{tab:best_fit}
 \end{table}

We see that values of $|\epsilon| \gtrsim 10$ are strongly disfavored by MINOS except
in the case of tiny active--sterile mixing angles. For such large values of $\epsilon$,
the new MSW resonance at $\Delta m_{41}^2 / (2E) \sim V_b$ lies within the MINOS
energy range $E < 50$~GeV and leads to a constraint $\sin^2 \theta_{24} \lesssim 10^{-3}$.
Such small mixing angles are, however, irrelevant
for possible explanations of MiniBooNE and other short-baseline anomalies.
The MINOS contours also show that in most of the mass range
$10^{-2}\ \text{eV}^2 \lesssim \Delta m_{41}^2 \lesssim 10^1\ \text{eV}^2$,
values of $\sin^2 \theta_{24} \gtrsim 0.01$ are excluded, with limits becoming much
stronger at large $\epsilon$.

Solar neutrinos also have some sensitivity to $\theta_{24}$, but limits on
$\epsilon$ vary a lot with $\sin^2 \theta_{24}$.  For intermediate values $0.01
\lesssim \sin^2 \theta_{24} \lesssim 0.1$, even values of $|\epsilon|$ as large
as $\text{few} \times 10^3$ are compatible with solar neutrino data.  For
$\epsilon > 0$, we notice that solar limits on $\epsilon$ are weakest at
$\sin^2 \theta_{24} \sim \text{few} \times 10^{-2}$. In this regime, the
additional neutrino disappearance due to nonzero $\theta_{14}$ and
$\theta_{24}$ is partially compensated by $V_b$-induced modifications to the
MSW resonance structure.  In particular, the 1--4 and 2--4 mixings imply that
above the solar MSW resonance, $\nu_1$--$\nu_2$ mixing is not as strongly
suppressed as in the standard case. This reduces the flavour transition
probability at energies above the resonance. Note that this effect is related
to a sterile neutrino-induced smearing of the \emph{atmospheric} resonance
(which at the center of the Sun lies at about 200~MeV) to the extent that it
has a small impact even at energies as low as $\sim 10$~MeV.  The effect is therefore
absent if the neutrino mass ordering is inverted so that the atmospheric
resonance lies in the anti-neutrino sector. We have checked that indeed the
limits on $\epsilon$ from solar neutrino experiments become somewhat weaker in
this case. For $\epsilon < 0$, the exclusion contours reveal an allowed
``island'' at $\epsilon \sim -10^3$.  In the parameter region corresponding to
these islands, the non-standard MSW resonance at $\Delta m_{41}^2 / 2 E \simeq
V_b$ mimics the effect of the standard solar resonance.  Also, in this parameter
region, the atmospheric MSW resonance---modified by the presence of the sterile
neutrinos---has a small impact.  Therefore, the ``islands'' move down by almost
an order of magnitude in $|\epsilon|$ if the neutrino mass ordering is
inverted.  The $\Delta m_{41}^2$-independent ``peninsula'' at $\epsilon \sim
-20$, is related to the appearance of the peak structure in $P_{\nu_e \to
\nu_e}$ which we discussed in section~\ref{sec:solar} and which is independent
of the mass ordering.

\begin{figure}
  \begin{center}
    \includegraphics[width=0.5\textwidth]{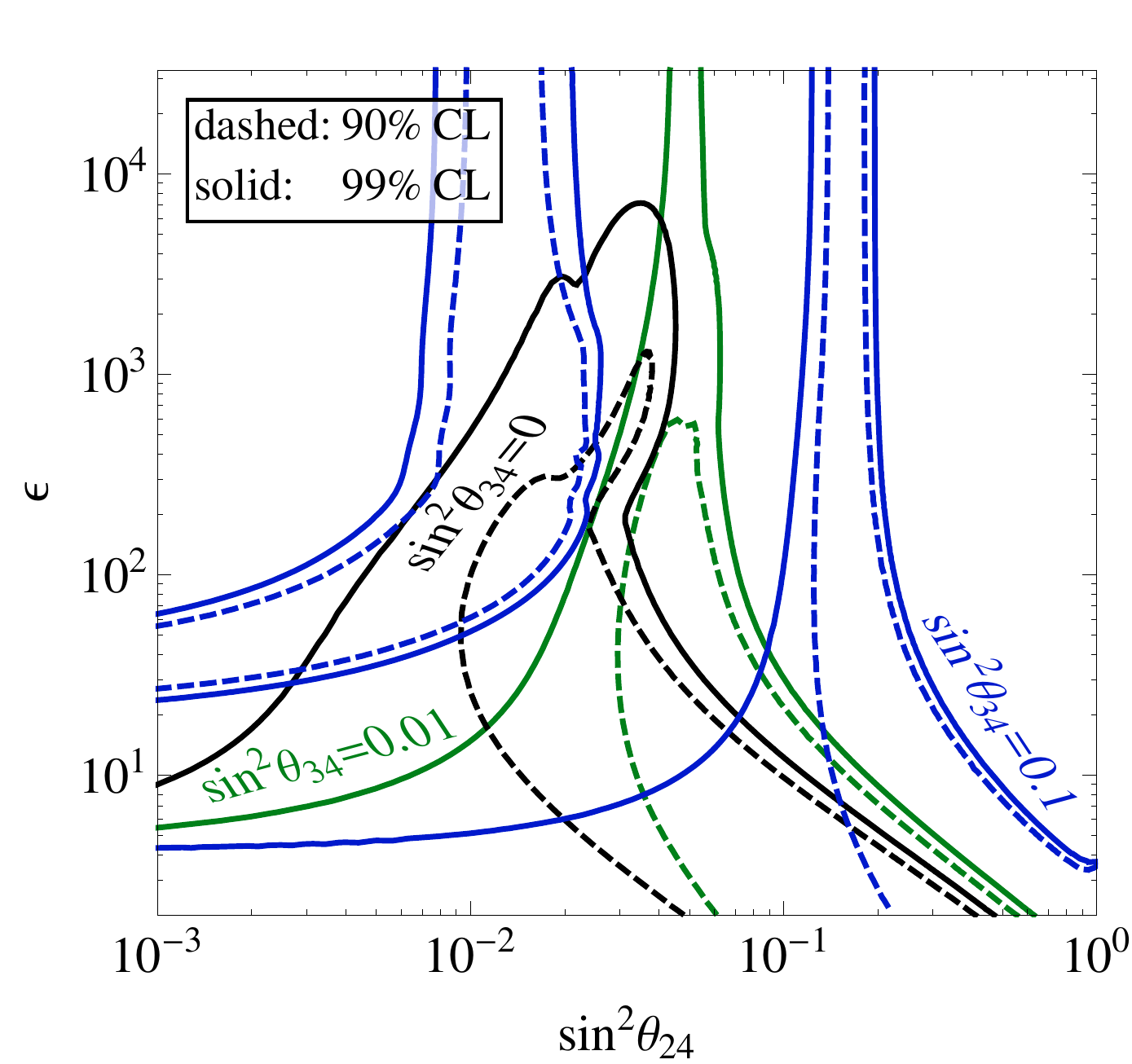}
  \end{center}
  \caption{Constraints on $\epsilon$ and $\sin^2 \theta_{24}$ from solar neutrinos for fixed
    $\sin^2 2\theta_{14} = 0.12$ (as motivated by
    the reactor and gallium anomalies), but for different values of $\theta_{34}$.
    The value of $\Delta m_{41}^2$ has been marginalized over in the range
    $10^{-2} \leq \Delta m_{41}^2 \leq 1.1 \cdot 10^1$.}
  \label{fig:solar-th34}
\end{figure}

The allowed parameter region for the measured appearance signal in MiniBooNE is
very similar to the one obtained in conventional sterile neutrino scenarios (see
for instance the analysis by the MiniBooNE collaboration themselves~\cite{AguilarArevalo:2012va})
with the exception
that for large matter potentials, the allowed region is expanded towards lower
$\sin^2 \theta_{24}$ and higher $\Delta m^2_{41}$. 

We now relax our assumption $\theta_{34} = 0$. The main sensitivity to $\theta_{34}$
is expected to come from solar neutrinos~\cite{Kopp:2013vaa} (and from MINOS neutral
current measurements, which we did not consider in this work, though). We
compare the solar neutrino limits in the $\sin^2 \theta_{24}$--$\epsilon$ plane for
different values of $\theta_{34}$ in figure~\ref{fig:solar-th34}, marginalizing over
the sterile neutrino mass in the range $10^{-2} \leq \Delta m_{41}^2 \leq 1.1 \cdot 10^1$.
We see that the constraints on $\epsilon$ become somewhat weaker if $\sin^2 \theta_{34} \sim 0.01$
and change significantly for larger values of $\sin^2 \theta_{34}$.  This implies that,
for large $\theta_{34}$, a scenario with strong non-standard matter potential can be
consistent with solar data and with MiniBooNE. Nevertheless, such a scenario would still
be ruled out by MINOS.

\begin{figure}
  \begin{center}
    \includegraphics[width=\textwidth]{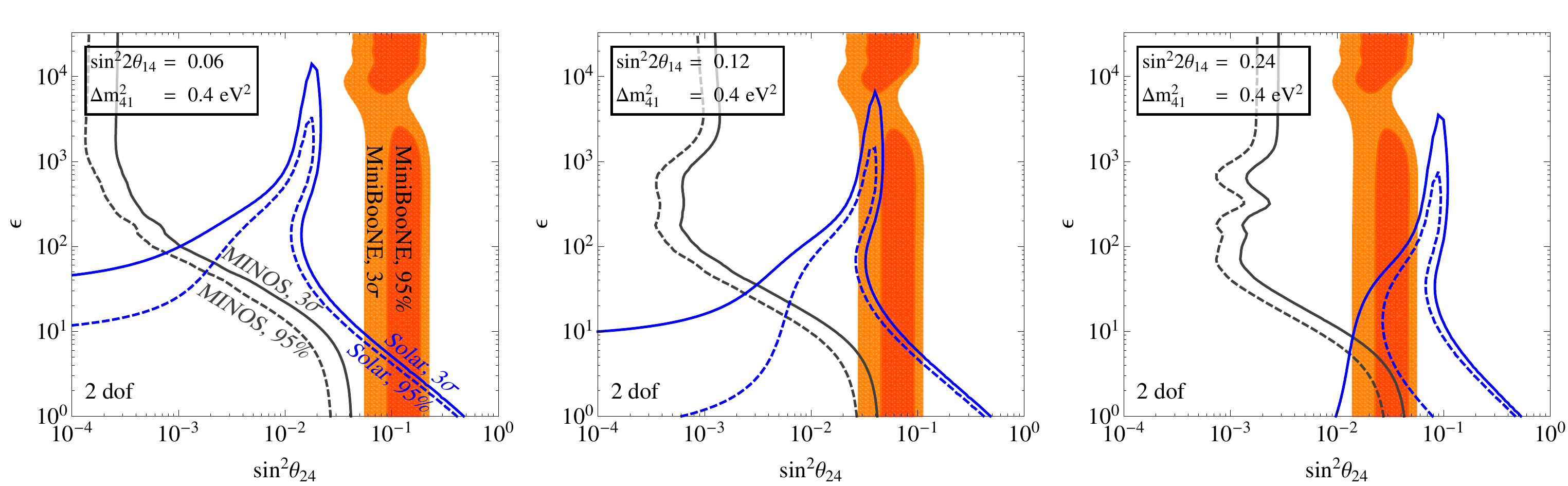}
  \end{center}
  \caption{The effect of varying $\theta_{14}$ on the constraints in the
    $\sin^2\theta_{24}$--$\epsilon$ plane. The plot in the center reproduces
    middle panel in the upper row of figure~\ref{fig:ResultsEPSpos}, while
    the left and right panels show similar constraints for smaller and larger $\theta_{14}$,
    respectively.}
  \label{fig:theta-14}
\end{figure}

Finally, let us also discuss the effect of choosing $\sin^2 2\theta_{14}$
different from the value 0.12 preferred by the reactor neutrino anomaly.  To
this end, we show in figure~\ref{fig:theta-14} how the constraints on
$\epsilon$ and $\theta_{24}$ for fixed $\Delta m_{41}^2$ are modified if
$\sin^2 2\theta_{14}$ is taken a factor of 2 smaller (left panel) or a factor
of 2 larger (right panel) than the preferred value.  We see that the MiniBooNE
preferred region, which is sensitive only to the combination $\sin^2
2\theta_{14} \sin^2 \theta_{24}$ is simply shifted by a factor of 2. Solar
limits are affected in a less trivial way and we find that at large
$\theta_{14}$, there is even a preference for nonzero $\theta_{24}$.  Note,
however, that the goodness of fit becomes slightly worse as $\theta_{14}$ is
increased: the minimum $\chi^2/\text{dof}$ is 0.8/3 for $\sin^2 2\theta_{14} =
0.6$ and 2.7/3 for $\sin^2 2\theta_{14} = 0.24$.  Finally, MINOS limits are
weakened if $\theta_{14}$ is large, especially at large $\epsilon$. This
happens because a larger mixing between $\nu_e$ and $\nu_b$ by unitarity
implies more $\nu_\mu$ disappearance.

\section{Conclusions}
\label{sec:conclusions}

To summarize, we have derived constraints on models with extended sterile
neutrino sectors that feature in particular a new gauge interaction between
sterile neutrinos and SM particles.  As a specific example, we have considered a
scenario in which eV-scale sterile neutrinos are charged under gauged baryon
number $U(1)_B$.  In principle, such interactions could be several orders of
magnitude stronger than SM weak interactions, so the
Mikheyev-Smirnov-Wolfenstein (MSW) potentials they generate could be
significantly larger than the matter potential in standard three-flavour
neutrino oscillations.

We have also computed approximate analytic expressions for the relevant
oscillation probabilities in matter, improving and extending the expressions
previously derived in~\cite{Karagiorgi:2012kw}.
We have then numerically analyzed data from the MINOS experiment, from solar
neutrino measurements and from MiniBooNE to show that new gauge interactions in
the sterile neutrino sector cannot be large unless the
active--sterile neutrino mixing is very small.  In particular, if the
ratio $\epsilon$ of the non-standard and standard matter potentials is larger
than $\sim 10$, MINOS excludes mixing angles down to $\sin^2 2\theta_{24} \sim
10^{-3}$. (This limit becomes stronger if $\theta_{14} = 0$.)

We conclude that sterile neutrino searches in oscillation experiments are
powerful tools to constrain certain models with hidden sector gauge
interactions. We also conclude that such models do not help to resolve the
tension in the global fit to short-baseline oscillation data.

Comparing to the interaction strength required for baryonic sterile neutrinos
to yield signals in dark matter detectors~\cite{Pospelov:2011ha, Harnik:2012ni,
Pospelov:2012gm, Pospelov:2013rha}, we conclude that in the case of eV scale
sterile neutrinos, baryonic interactions cannot be large enough to be
observable in the current generation of experiments. On the other hand,
interesting signals may still be possible in future ton-scale experiments.

\section*{Acknowledgments}

It is a pleasure to thank Janet Conrad and Georgia Karagiorgi for
very helpful discussions.

\bibliographystyle{apsrev}
\bibliography{./gauged-steriles}

\end{document}